# Large-transverse-momentum processes: the ISR as a gluon collider

*P. Darriulat*

**Abstract**
It is argued that, contrary to what is often said, large-transverse-momentum hadronic processes studied at the ISR have made a significant contribution to the understanding of the strong interaction and, in particular, to the development of quantum chromodynamics. In their unique role as a gluon collider the ISR have provided information that no other accelerator could have directly offered. They allowed one to probe high values of the centre-of-mass energy that were not available to fixed-target experiments. The latter, however, were more flexible and, together, they allowed for powerful explorations of the hadron structure and of the relevant dynamics in sectors such as inclusive particle production, direct photon production, and jet structure studies. It remains true that, rightly so, the ISR will be mostly remembered as the founders of a lineage that includes the proton–antiproton colliders and, today, the LHC.

## 1      Introduction

It so happens that the lifetime of the Intersecting Storage Rings (ISR), roughly speaking the 1970s, coincides with a giant leap in our understanding of particle physics. However, it is honest to say that, to first order, there is no causal relation between the two. Yet, those of us who have worked at the ISR remember these times with the conviction that we were not merely spectators of the ongoing progress, but also — admittedly modest — actors. The ISR contribution, it seems to us, is too often unjustly forgotten in the accounts that are commonly given of the progress of particle physics during this period. In the present article, I try to present arguments of relevance to this issue in what I hope to be as neutral and unbiased a way as possible. I restrict the scope of my presentation to large-transverse-momentum processes, or equivalently to the probing of the proton structure at short distances. This, however, is not much of a limitation, as the ISR did not significantly contribute to the progress achieved in the weak sector.

Anyone trying to reconstruct history is prompt to learn that each individual has his own vision of what has happened in the past and that history can merely be an attempt at collecting all such visions into as coherent as possible a story. As David Gross reminds us [1], quoting Emerson, *"There is properly no history; only biography"*. In physics, this is particularly true when discoveries and new ideas occur at a rapid pace, as was the case in the 1970s. Each of us remembers a seminar, a discussion at coffee, the reading of a particular article, or another event of this kind as a milestone in his own understanding of the new ideas. For most of us, it has no incidence on the history of physics: I understood superconductivity 40 years after BCS and general relativity 90 years after Einstein... But for those having played a major role in the blooming of the new ideas, it has. For example, reading accounts by Steve Weinberg [2], David Gross [1], Gerard 't Hooft [3] or Jerry Friedman [4] of how they remember this period is particularly instructive in this respect.

The same kind of disparity that exists between the visions of different individuals also occurs between the visions of different communities. In particular, during the 1970s, the $e^+$-$e^-$ community, the neutrino community, the fixed-target community, and the ISR community have all had quite different perceptions of the progress that was being achieved. It is therefore useful to recall briefly the main events in this period.

## 2      The main milestones

When Vicky Weisskopf, in December 1965, in his last Council session as Director-General obtained approval for the construction of the ISR, there was no specific physics issue at stake, which the machine was supposed to address; its only justification was to explore the *terra incognita* of higher-centre-of-mass-energy collisions (to my knowledge, since then, all new machines have been proposed and approved with a specific physics question in mind, which they were supposed to answer). The strong interaction was perceived as a complete mystery. The eightfold way, today understood as the approximate *SU(3)* flavour symmetry associated with interchanges of *u, d* and *s* quarks, was not believed to have significant consequences in the dynamics of the strong interaction. The fact that no free quark had been found in spite of intensive searches, and that states such as $\Delta^{++}$, with spin-parity $3/2^+$, could not be made of three identical spin-½ *u* quarks without violating Fermi statistics, were discouraging such interpretations.

The first hint to the contrary came in 1968–1969 at SLAC [4] with the discovery of an important continuum in the deep-inelastic region of electron proton scattering. The 2-mile linear accelerator had started operation the preceding year and the experimental programme, using large spectrometers, extended over several years. From the very beginning, experimenters and theorists were in close contact, feeding each other with new data and new ideas, starting with Bjorken's ideas on scaling [5] and Feynman's ideas on partons [6], both early advocates of a proton structure consisting of point-like constituents. However, one had to wait until 1972 for the case for a quark model to become strong: by then, scaling had been established; the measurement of a small *R* value (the ratio of the absorption cross-sections of transverse and longitudinal virtual photons) had eliminated competitors such as the then popular Vector Dominance Model; deuterium data had been collected allowing for a comparison between the proton and neutron structure functions; a number of sum rules had been tested; evidence for the quarks to carry but a part of the proton longitudinal momentum had been obtained; the first neutrino deep-inelastic data from Gargamelle had become available [7]. By the end of 1972, the way was traced for Gross, Wilczek, and Politzer [8] to conceive the idea of asymptotic freedom and its corollary, infrared slavery, explaining why one could not see free quarks. By the end of 1973, the connection with non-Abelian gauge theories had been established and the "advantages of the colour-octet gluon picture", including the solution of the Fermi statistics puzzle, had been presented by Fritzsch, Gell-Mann, and Leutwyler [9]. QCD was born and, by 1974, was starting to be accepted by the whole community as ***the*** theory of the strong interaction. It took another three to four years for it to come of age.

By mid 1972, SPEAR, the Stanford electron–positron collider, had begun operation. In November 1974, it shook the physics community with what has since been referred to as a Revolution: the discovery of the $\Psi$ going hand in hand with the simultaneous discovery of the *J* at Brookhaven. It immediately exploited its ability to produce pure quark–antiquark final states to measure the number of colours. However, there were so many things happening in the newly available energy domain (opening of the naked charm channels, crowded charmonium spectroscopy, production of the $\tau$ lepton) that it took some time to disentangle their effects and to understand what was going on. By the end of the decade, scaling violations had been studied both in neutrino interactions and in electron–proton annihilations (DORIS had started operation in Hamburg two years after SPEAR). QCD had reached maturity and the only puzzling questions that remained unanswered, the absence of a CP-violating phase and our inability to handle the theory at large distances, are still with us today.

## 3      What about the ISR?

The above account of the progress of particle physics in the 1970s, while following the standard folklore, does not even mention the name of the ISR. I remember having asked David Gross whether he was aware of the results obtained at the ISR and whether they had an impact on the development of QCD. His answer [10] was: *"Every one was aware of the qualitative phenomena observed in hadronic physics at large $p_T$, which were totally consistent with simple scattering ideas and parton* *model ideas [...] The tests were not as clean as in deep inelastic scattering, the analysis was more difficult and deep inelastic scattering was much cleaner in the beginning of perturbative QCD [...] Parton ideas did not test QCD at all, they simply tested the idea that there were point-like constituents but not the dynamics."* Alvaro de Rujula, who witnessed from Boston "the maiden years of QCD", being asked the same question, simply answered [10]: *"I do not know the answer to this question, I am not an historian"*. Such answers illustrate well the way in which the ISR were generally perceived: a collider that was shooting Swiss watches against each other, as Feynman once jokingly described. Yet, some theorists followed closely what the ISR were producing; paradoxically, Feynman was one of them, Bjorken was another.

David Gross could have returned the question to me: *"How aware were you, the ISR community, of the experimental progress at SLAC and of the new ideas in theory?"* The first name that comes to mind in answer to this question is that of Maurice Jacob. Maurice had spent a sabbatical at Stanford where, together with Sam Berman, he had written a seminal paper on point-like constituents and large-transverse-momentum production [11]. Back at CERN, he organized a lively series of discussions between ISR experimenters and theorists that proved to be extremely successful in permeating our community with the progress in deep-inelastic scattering and, later, in electron–positron collisions. At that time, our community was small enough to fit in the ISR auditorium. Maurice was gifted with an unusual talent to make theoretical ideas accessible to us. We all remember these seminars as a most profitable experience that brought coherence and unity in our community. For this reason, it makes sense to talk about a common ISR culture. In particular, by 1972, we were aware of the basic parton ideas and of the picture of large-transverse-momentum production

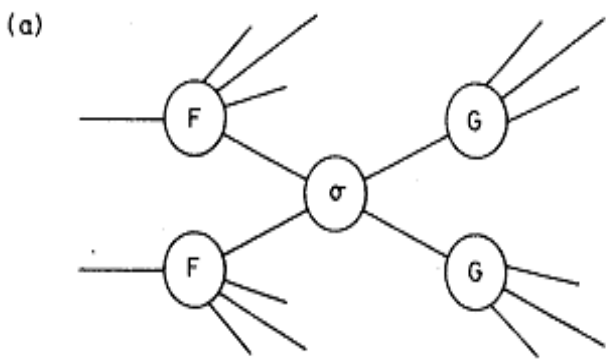

**Fig. 1:** Parton model picture of high-$p_T$ hadron interactions. One parton of each of the incident hadrons (structure function F) experiences a binary collision ($\sigma$) and the outcoming partons fragment into hadrons (fragmentation function G)

factorized in three steps (Fig. 1): singling out a parton in each proton, making them interact (how, was not clear) in a binary collision and letting the final-state partons fragment into hadrons. There were a few papers [6, 11–16] in support of such a picture which most of us had read and which were our basic reference. Yet, in these early days, there was a typical delay of at least six months between SLAC and us for a new idea to be digested. There was even more delay, for most of us, to digest the more subtle development of non-Abelian gauge theories: we only knew about it from our theorist friends.

Table 1 lists leading-order diagrams involving quarks or gluons. A simple glance at it illustrates the originality of the ISR: gluons contribute to leading order. In electron–proton annihilations and deep-inelastic scattering, gluons contribute to next-to-leading order only, in the form of radiative corrections associated with a bremsstrahlung gluon radiated from a quark line. This does not mean that such gluon contributions are unimportant: the scaling violations which they induce have been one of the most powerful tool in the development of our understanding of QCD. But, at the ISR, gluons not only contribute to leading order but indeed dominate the scene: in the low $x$ regime characteristic of the ISR, collisions involving gluons, either gluon–gluon or quark–gluon, account for most of the high-$p_T$ cross-section. Gluon interactions being a privileged domain of the ISR, and gluons having been the last component of the theory to be understood and digested, it seems difficult to argue that the ISR have played but a minor role. The more so when one considers that the ISR had exclusive access to the three- and four-gluon vertices, which are a specific expression of QCD as a non-Abelian gauge theory.

**Table 1:** Leading order processes involving quarks or gluons

| | | | |
|---|---|---|---|
| | Electron–positron annihilations | | |
| 1 | | $e^+e^- ><  q^+q^-$ | $\alpha^2 G^2$ |
| | Deep-inelastic electron scattering | | |
| 2 | | $eq]\gamma[eq$ | $\alpha^2 FG$ |
| | Deep-inelastic neutrino scattering | | |
| 3 | Neutral currents | $vq]Z[vq$ | $\alpha_n^2 FG$ |
| 4 | Charged currents | $vq]W[lq$ | $\alpha_{ch}^2 FG$ |
| | Proton–proton collisions (ISR) | | |
| 5 | Drell–Yan | $q^+q^- ><  l^+l^-$ | $\alpha^2 F^2$ |
| 6 | Direct photons | $q^+q^-]q[\gamma g$ | $\alpha\alpha_s F^2 G$ |
| 7 | | $qg]q[\gamma q$ | |
| 8 | Large $p_T$ hadrons | $qq]g[qq$ | $\alpha_s^2 F^2 G^2$ |
| 9 | | $qq]q[gg$ | |
| 10 | | $q^+q^- > g <  gg$ | |
| 11 | | $q^+q^- > g <  q^+q^-$ | |
| 12 | | $qg]q[qg$ | |
| 13 | | $qg]g[qg$ | |
| 14 | | $qg > q <  qg$ | |
| 15 | | $gg > g <  q^+q^-$ | |
| 16 | | $gg > g <  gg$ | |
| 17 | | $gg]q[qq$ | |
| 18 | | $gg]g[gg$ | |
| 19 | | $gg > <  gg$ | |

We note $s$ channel exchange as $><$ and $t$ channel exchange as $][$. When necessary, quarks are written $q^+$ and antiquarks $q^-$. The last column gives the coupling constants, the number of structure functions ($F$), and the number of fragmentation functions ($G$) taking part in the cross section. The couplings are written $\alpha_n$ for $\alpha/(\sin\theta_W \cos\theta_W)^2$ and $\alpha_{ch}$ for $\alpha/\sin\theta_W^2$ with $\theta_W$ being the Weinberg angle. Processes involving gluons in the initial state are shaded.

## 4 Large transverse momentum: inclusive production data

In 1972–1973, three ISR teams [17–19] announced the observation of an unexpectedly copious pion yield at large transverse momenta (Fig. 2), orders of magnitude above a (traditionally called naïve) extrapolation of the exponential distribution observed at low-$p_T$ values, $\sim\exp(-6p_T)$. "Unexpectedly" is an understatement. The whole ISR experimental programme had been designed under the assumption that all hadrons would be forward-produced. The best illustration was the Split Field Magnet, meant to be the general multipurpose detector at the ISR. No experiment was equipped with very large solid angle good-quality detectors at large angle. This first discovery was opening the ISR to the study of large-transverse-momentum production and was providing a new probe of the proton structure at short distances. That was the good side of it. But it also had a bad side: the background that had been anticipated in the search for new particles had been strongly underestimated and such searches were now becoming much more difficult than had been hoped for.

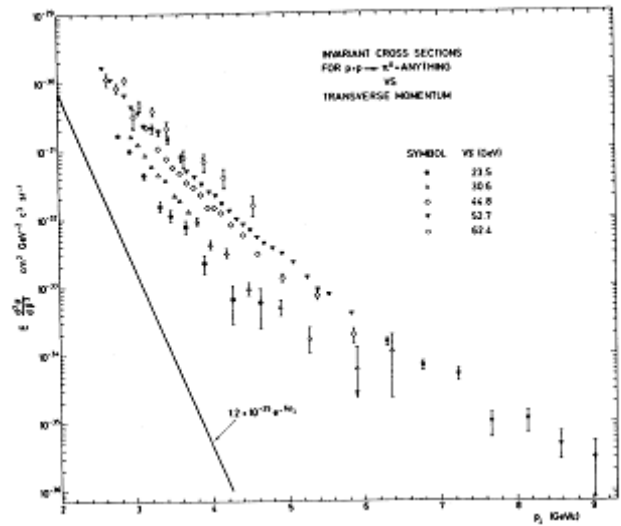

**Fig. 2:** Early inclusive $\pi^0$ cross-section [20] giving evidence for copious production at high $p_T$ well above the exponential extrapolation of lower energy data

Bjorken scaling was found to apply, in support of the parton picture, but the index of the $p_T$ power law was twice as high as the value expected from point-like constituents, 8 rather than 4. Precisely, the $\pi^0$ inclusive invariant cross-section was of the form $p_T^{-n} \exp(-kx_T)$ where $x_T = 2p_T/\sqrt{s}$, $n = 8.24 \pm 0.05$ and $k = 26.1 \pm 0.5$. The impact of this result was quite strong and brought into fashion the so-called constituent interchange model [20]. The idea was to include mesons in addition to quarks among the parton constituents of protons: deep-inelastic scattering would be blind to such mesons because of their form factor but hadron interactions would allow for quark rearrangements such as $\pi^+ + d \to \pi^0 + u$. At large values of $x_T$, the cross section was then predicted to be of the form $p_T^{-2(n-2)}(1-x_T)^{2m-1}$ where $n$ stands for the number of "active quark lines" taking part in the hard scattering and $m$ stands for the number of "passive" quark lines wasting momentum in the transitions between hadrons and quarks. The model, that correctly predicted the power 8 measured at the ISR, had many successes but did not stand the competition with early QCD models that were starting to be developed. Such an example is illustrated in Fig. 3, giving evidence for important quark–gluon and gluon–gluon contributions [21] beside the quark–quark term. By then, the inclusive production of charged pions, kaons, protons, and antiprotons as well as $\eta$ mesons had been studied at the ISR, and at Fermilab where a $\pi^-$ beam had also been used, providing decisive evidence in favour of QCD. It was then understood that the $p_T$ power law was indeed evolving to $p_T^{-4}$ at high values of $x_T$, which, however, were only accessible, in practice, to larger-centre-of-mass-energy collisions. The successes of the constituent interchange models were then relegated to the rank of "higher twist corrections" to the leading-order perturbative regime.

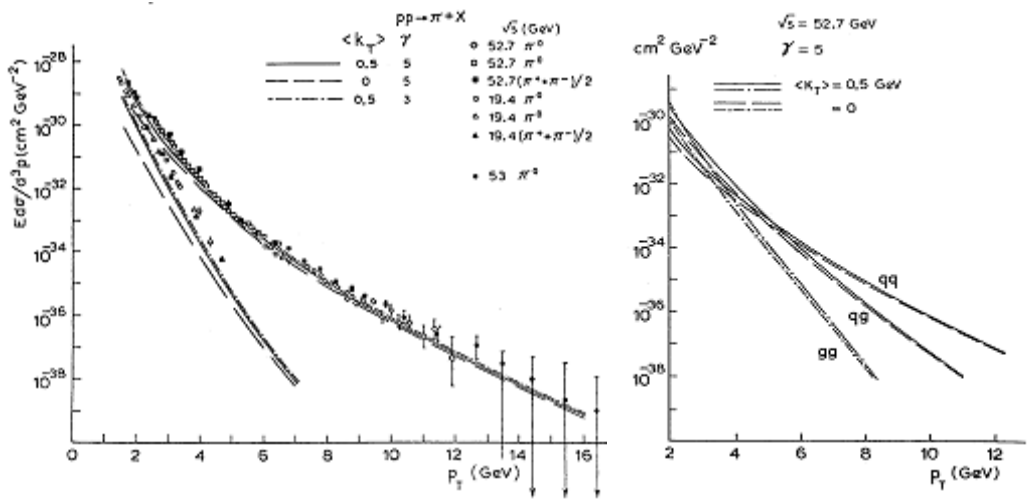

**Fig. 3:** A typical QCD fit [21] to inclusive pion data (left) and the relative contributions of quark–quark, quark–gluon and gluon–gluon diagrams (right)

Between 1973 and 1978, inclusive high-$p_T$ single-hadron production in hadron collisions had given exclusive contributions to the establishment of QCD as the theory of the strong interaction in a domain where other experiments — deep-inelastic scattering and electron–positron annihilations — could not contribute: that of short-distance collisions involving gluons to leading order of the perturbative expansion. In this domain, the data collected at the CERN ISR — at the higher-centre-of-mass energies — and at Fermilab — with a variety of beams and targets — nicely complemented each other. As the results were confirming the validity of QCD, and as there were so many important events happening elsewhere in physics, people tended to neglect or forget these important contributions.

# 5 Event structure and jets

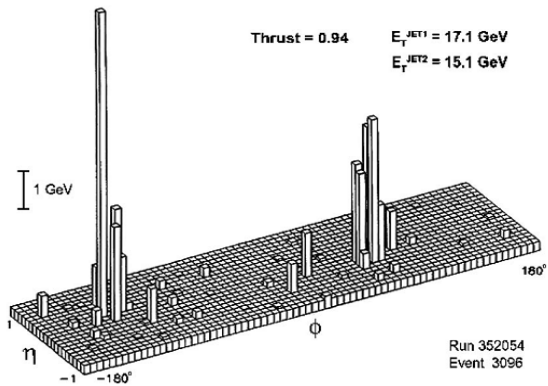

**Fig. 4:** A lego plot from the AFS experiment showing the two-jet structure that dominates at larger transverse energies. (from Ref. [23])

The early evidence in favour of the parton picture encouraged studies of the global event structure and, in particular, experiments aiming at the detection of the hadron jets into which the hard-scattered partons were supposed to fragment. Unfortunately, none of the existing ISR detectors was matched to the task. In March 1975, a large magnetic detector serving precisely this purpose had been proposed to the ISR Committee by a collaboration of British, Scandinavian, and US physicists but had been rejected in October of the same year. The proposal had been reiterated with various amendments. It was enjoying the support of the ISR community, of a Working Party that had been appointed to assess "the need for a new magnetic facility at the ISR", with Nino Zichichi in the chair, and of the ISR Committee (69th meeting, November 10th, 1976). It was definitively turned down two weeks later by the Research Board. Meanwhile, step by step, the existing ISR experiments had upgraded their set-ups as well as they could but one had to wait until 1982, with the Axial Field Spectrometer in I8 and the Superconducting Solenoid in I1 to see detectors having large calorimeter coverage (electromagnetic and hadronic for the former but only electromagnetic for the latter). When the ISR closed down in 1984, a rich set of important results had been obtained by these two groups [22], with two-jet events (Fig. 4) dominating the scene for transverse energies in excess of 35 GeV [23]; but the CERN proton–antiproton collider, which had published its first jets in 1982 [24], had already taken the limelight away from the ISR.

There is no doubt that the lack of proper instrumentation has been a major handicap for the ISR in their contribution to the physics of hard collisions. More support from the management would probably have made it possible to gain two precious years. Retrospectively, it is difficult to estimate how much of a negative impact the approval of a new large facility at the ISR would have had on the high-priority CERN programmes, LEP and the proton–antiproton collider. There is no doubt that these were the machines where quark and gluon jets could be studied in optimal conditions: in comparison, the ISR were quite marginal. Moreover, the ISR beam geometry, with a crossing angle of 15° and the need for large vacuum chambers, was making the design of a 4π detector difficult. Seen from today, thirty years later, our frustration was certainly understandable and legitimate, but the decision of the management sounds now more reasonable than it then did.

Between 1973 and 1978, several ISR experiments had completed studies of the event structure and the evidence for hard jets in the final state, already clear in 1976 [25], had become very strong. Figure 5 shows the longitudinal phase-space density of charged particles produced in a hard-scattering collision. It is an average of data collected by the British–French Collaboration using a charged-particle trigger at 90° and momentum analysing in the Split Field Magnet the charged particles produced in association. Particle densities are normalized to those obtained in minimum-bias collisions. Particle densities are normalized to those obtained in minimum-bias collisions. Several features are visible: diffraction is suppressed at large rapidities, a 'same-side' jet is present alongside the trigger and 'away-side jets', at opposite azimuth to the trigger, cover a broad rapidity range.

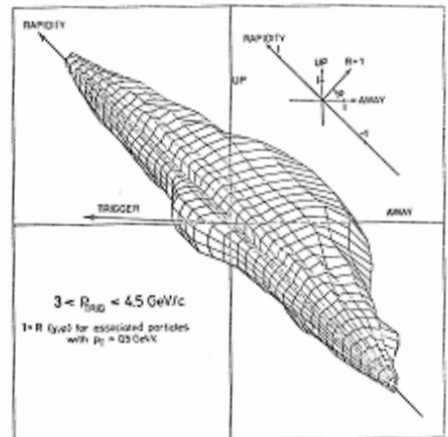

**Fig. 5:** Longitudinal phase-space density (relative to minimum-bias events) associated with a single particle trigger at 90° (see text)

A difficulty inherent to the study of hard hadron collisions is the presence of a so-called 'underlying event' which contains the fragments of the spectator partons that do not take part in the hard collision. This is at variance with electron–positron annihilations where all hadrons are fragments of the hard scattered partons and, to a lesser extent, with deep-inelastic scattering where most of the information is carried by the structure functions. It implies a transverse momentum threshold, half a GeV to one GeV, below which a particle cannot be unambiguously identified as being a fragment of a hard scattered parton. At ISR energies, it is a serious limitation.

A second difficulty, resulting from the lack of proper calorimeter coverage in the first decade of ISR operation, was the so-called 'trigger bias'. Since the hard parton scattering cross-section has a much steeper $p_T$ dependence than has the fragmentation process, it is very likely for a particle of a given $p_T$ to be the leading fragment of a rather soft jet. This distortion of the 'same-side' jet fragmentation creates an asymmetry between it and the 'away-side' jet, which makes it more difficult to compare their properties. For this reason, an ideal experiment should trigger on the total transverse energy $E_T$ using calorimetric devices. Numerous studies of the 'same-side' correlations have been performed at the ISR, establishing early that they were not the result of resonance production but of a jet fragmentation characterized by a limited transverse momentum around the jet axis.

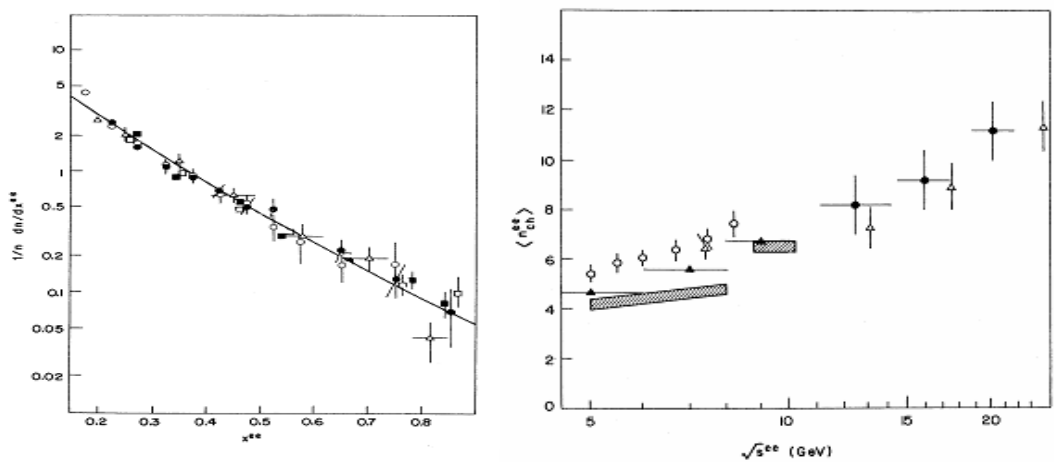

**Fig. 6:** Left: Jet fragmentation functions measured in different processes (triangles are for neutrino deep-inelastic, circles for high-$p_T$ hadronic interactions at the ISR and the solid line for $e^+e^-$ annihilations). Right: Mean charge multiplicity of hadron jets as a function of the equivalent $e^+e^-$ energy as measured at SPEAR and DORIS (cross-hatched rectangles), at PETRA (open triangles), in neutrino deep-inelastic scattering (full triangles) and in high-$p_T$ hadronic interactions at the ISR (circles)

Evidence for an excess of particles at opposite azimuth to the trigger had been obtained very early and it had soon been recognized that it was due to a collimated jet produced at a rapidity which was different from event to event. The away-side jet multiplicity could then be measured and compared to that of quark jets observed in deep inelastic and electron–positron annihilations (Fig. 6 right). ISR jets being dominantly gluon jets, one could expect to see a difference but the $p_T$ range accessible to the ISR was still too low to reveal significant differences in the fragmentation functions of quark and gluon jets (Fig. 6 left).

In electron–positron collisions, the first evidence for quark jets came from SPEAR in 1975 [26] and the first evidence for gluon jets came from PETRA in 1979–1980 [27]. The former were 4 GeV quark jets, PETRA's gluon jets were typically 6 GeV, ISR jets — mostly gluon jets — were at least 10 GeV. The $e^+e^-$ data were analysed in terms of event shapes: sphericity, oblateness, thrust, triplicity, etc. There was no doubt that, without any theoretical preconception, the evidence for ISR jets was stronger than the evidence for quark jets at SPEAR in 1975 and the evidence for gluon jets at PETRA in 1979–1980; the ISR physicists who studied large-transverse-momentum production were rightly feeling frustrated with the relative lack of public recognition given to their data compared with the enthusiasm generated by the SPEAR and PETRA results. The worst sceptics were to be found in the fixed-target community where too low values of the centre-of-mass energy prevented jets from being revealed. There were exceptions, however. I remember Walter Selove spending the Summer months at CERN and scanning with us our streamer chamber data collected with a high-$p_T$ $\pi^0$ trigger at 90°: each time he would see some kind of a jet, he would exult and copy its configuration in a notebook.

Part of the imbalance in the reception given to ISR data compared with SPEAR and PETRA data was subjective: the analysis of ISR data was too complicated, which for many meant "was not clean". But, one must recognize that a good part was objective. First because the SPEAR and PETRA detectors were better fitted to these kinds of studies and second, more importantly, because good physics is done with, rather than without, theoretical preconception. In the SPEAR case, the beauty of their results came from two important features which gave strong support to the quark jet hypothesis: the azimuthal distribution of the jet axis displayed the behaviour expected from the known beam polarization and its polar angle distribution obeyed the $1 + \cos^2\theta$ law expected in the case of spin -½ partons. In the PETRA case, by mid-1980, all four experiments had presented clear evidence for gluon bremsstrahlung, including convincing comparisons with QCD predictions.

At the ISR, the complexity of the physics processes at stake was undoubtedly much larger than at electron–positron colliders, making it difficult to devise decisive QCD tests independent from what had been learned at other accelerators. But, once again, ISR data were exploring elementary processes which were not accessible to other accelerators and were shown to nicely fit in a coherent QCD picture embedding deep-inelastic as well as $e^+e^-$ annihilation results. This was clearly an independent and essential contribution to the validation of QCD.

## 6    Photons and leptons

Leptons were produced at the ISR either as decay products of other particles or as a continuum of opposite-charge pairs coupled to a quark–antiquark pair in the initial state via a virtual photon in the $s$ channel, the so-called Drell–Yan process. In the first half of the decade, the $e/\pi$ ratio had been measured by several experiments to be of the order of $10^{-4}$ over a broad range of transverse momenta and was understood as being the result of a 'cocktail' of different sources, including, among others, open charm and charmonium. By the end of the decade, the $J/\Psi$ and the $\Upsilon$ had been detected and their production cross-section had been measured. Moreover, a clear evidence for $D$ production [28] had been obtained at the Split Field Magnet — for the first time in hadron interactions. Dilepton masses up to 20 GeV have been ultimately studied, giving evidence for strong next-to-leading-order corrections to the Drell–Yan leading-order diagram.

The production of direct photons was soon recognized to be a particularly simple process: its comparison with QCD predictions could be expected to be instructive. It proceeds either by a quark–antiquark pair in the initial state radiating a photon and a gluon in the final state or by a Compton-like interaction between a quark and a gluon producing a quark and a photon. In both cases, the photon is produced alone, without high-$p_T$ companions, and its transverse momentum is balanced by a hadron jet. At the ISR, the Compton diagram dominates: the study of direct photon production should provide information on the gluon structure function as well as a measurement of $\alpha_s$, the quark fragmentation being borrowed from $e^+e^-$ data. In the first half of the decade, pioneering measurements established the existence of a signal and identified backgrounds, the main source being $\pi^0$ and $\eta$ decays sending one of the two decay photons alongside their own momentum. At the end of the decade, clear signals were observed [29, 30] and a series of measurements followed, which, together with fixed-target data, provided a very successful laboratory for QCD (Fig. 7). Once again, hadronic interactions, both on fixed-target machines and at the ISR, had made use of their unique ability to study gluon collisions and to give essential contributions to the study of the strong interaction in the QCD perturbative regime [31].

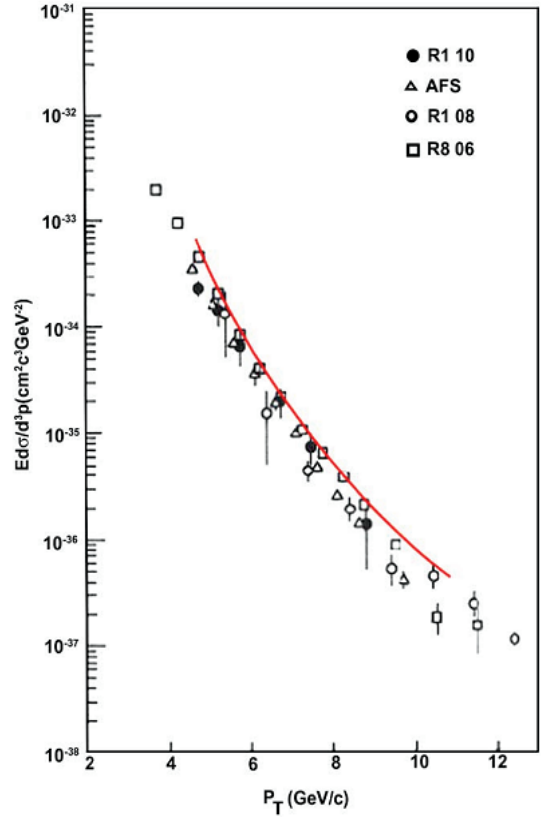

**Fig. 7:** Experimental invariant cross-sections for direct photon production (compilation by L. Camilleri) are compared with a next to leading order QCD calculation (by P. Aurenche and M. Werlen), from Ref 24.

## 7    The ISR legacy

I hope that this brief review of ISR contributions to the new physics that was born in the 1970s, and specifically to QCD becoming the theory of the strong interaction, has convinced the reader that they were more than a mere test of the idea that there were point-like constituents inside the proton. Together with hard hadron interactions on fixed-target machines, they made optimal use of their exclusive property to study the gluon sector of QCD to leading order. The ISR had the privilege of a higher centre-of-mass energy, fixed-target machines had the privilege of versatility, their respective virtues nicely complemented each other. Many factors have contributed to the relative lack of recognition which has been given to ISR physics results: the absence, for many years, of detectors optimized for the study of hard processes, the fact that the weak sector, which during the decade was the scene of as big a revolution as the strong sector, was completely absent from the ISR landscape and, may be most importantly, the fact that hard hadron collisions imply complex processes which may seem 'dirty' to those who do not make the effort to study them in detail.

We, who worked at the ISR, tend not to attach much importance to this relative lack of recognition because for us, their main legacy has been to have taught us how to make optimal use of the proton–antiproton collider, which was soon to come up. They had given us a vision of the new physics and of the methods to be used for its study which turned out to be extremely profitable. They had played a seminal role in the conception of the proton–antiproton collider experiments, they were the first hadron collider ever built in the world, they were the machine where a generation of physicists learned how to design experiments on hadron colliders. We tend to see the ISR and the proton–antiproton colliders, both at CERN and at the Tevatron, as a lineage, father and sons, the success of the latter being indissociable from the achievements of the former.

We were young then, this may be another reason why we remember these times with affection. With the LHC coming up, the lineage has now extended to a third generation and we look at the future with the eyes of grandparents, full of tenderness and admiration for their grandson whom we wish fame and glory.

**Acknowledgements**

It is a pleasure to thank the Director-General and the organizers of the Colloquium for their very successful initiative and for having invited me to give this brief recollection. I am deeply grateful to Luigi Di Lella, Antonino Zichichi and Giorgio Bellettini for pertinent comments on the manuscript and at the end of the oral presentation.